\documentclass[aps,prx,twocolumn,floatfix,superscriptaddress,showpacs,showkeys]{revtex4-1}
\usepackage{epsf,amsmath,amssymb,verbatim,color,multirow,pifont,graphicx,hyperref,cleveref,tabularx,xcolor,soul}

\crefname{equation}{Eq.}{Eqs.}
\crefname{subsection}{Sec.}{Secs.}
\crefname{figure}{Fig.}{Figs.}
\crefname{table}{Table}{Tables.}

\newcommand{\nbar}{\bar{\nu}}
\newcommand{\nbara}{\nbar_a}

\newcommand{\psa}{p_{s}}

\newcommand{\asd}{avalanche size distribution}

\begin{document}

\title{Critical behavior of $k$-core percolation: Numerical studies}
\author{Deokjae Lee}
\affiliation{CCSS,  CTP and Department of Physics and Astronomy, Seoul National University, Seoul 08826, Korea}
\author{Minjae Jo}
\affiliation{CCSS,  CTP and  Department of Physics and Astronomy, Seoul National University, Seoul 08826, Korea}
\author{B. Kahng}
\email{bkahng@snu.ac.kr}
\affiliation{CCSS,  CTP and  Department of Physics and Astronomy, Seoul National University, Seoul 08826, Korea}
\date{\today}

\begin{abstract}
$k$-Core percolation has served as a paradigmatic model of discontinuous percolation for a long time. Recently it was revealed that the order parameter of $k$-core percolation of random networks additionally exhibits critical behavior. Thus $k$-core percolation exhibits a hybrid phase transition. Unlike the critical behaviors of ordinary percolation that are well understood, those of hybrid percolation transitions have not been thoroughly understood yet. Here, we investigate the critical behavior of $k$-core percolation of Erd\H{o}s-R\'enyi networks. We find numerically that the fluctuations of the order parameter and the mean avalanche size diverge in different ways. Thus, we classify the critical exponents into two types: those associated with the order parameter and those with finite avalanches. The conventional scaling relations hold within each set, however, these two critical exponents are coupled. Finally we discuss some universal features of the critical behaviors of $k$-core percolation and the cascade failure model on multiplex networks. 
\end{abstract}


\maketitle

\section{Introduction}
Recently hybrid phase transitions (HPTs) containing natures of both first-order and second-order phase transitions have drawn much attention ~\cite{kcore1,jamming, dodds,pazo,kcore2,kcore3, baxter,mendes_sync,mukamel,zhou,grassberger_natphys,rermodel,mcc_lee}. The order parameter changes discontinuously at a transition point and exhibits a critical behavior in the vicinity of the transition point, but some physical quantities such as the susceptibility and correlation length diverge at a transition point. 
However, a thorough study is yet to be conducted to determine if the conventional scaling relations among the critical exponents hold for the critical behavior of HPTs  as they do for second-order phase transitions.  In a recent paper \cite{mcc_lee}, we studied the scaling relations in the HPT of the cascading failure (CF) model on multiplex networks~\cite{buldyrev}. Unlike ordinary percolation, the HPT exhibits two different critical behaviors: divergences of the fluctuations of the order parameter and mean avalanche size of finite avalanches at a transition point in different ways. Thus, two sets of critical exponents are required, those associated with the order parameter and those with finite avalanches. We found that conventional scaling relations hold only among the critical exponents associated with the order parameter and do  partially for finite avalanches. Our studies were helpful in understanding the critical behavior and thus establishing a theoretical framework for the hybrid percolation transition in pruning processes. 

In this paper, we extend our previous study for the CF model to $k$-core percolation on single-layered Erd\H{o}s-R\'enyi (ER) networks. 
The critical behavior of $k$-core percolation is analytically accessible more easily than that of the CF model. Thus, some of the critical exponents have been obtained analytically using the local tree approximation~\cite{kcore2,kcore3,kcore_prx}. Yet, a complete set of critical exponents for $k$-core percolation, particularly those associated with hyperscaling relations have not been determined even numerically. Thus, the aim of this paper is to determine all critical exponent values of $k$-core percolation.  We find that even though the detailed dynamics of the CF model on multiplex networks and $k$-core percolation are different, the critical behaviors of each system share some universal features. For instance, the hyperscaling relations hold within the critical exponents associated with the order parameter, and there exists a relation between the exponents of the order parameter and the mean size of finite avalanches. Most critical exponents are the same.

The paper is organized as follows:  In section II, we introduce $k$-core percolation on ER random networks and show the critical behavior of hybrid phase transition that the $k$-core percolation exhibits. Moreover, we introduce two sets of exponents representing the critical behaviors. In section III,  we present numerical results for the critical exponents associated with the order parameter and finite avalanches, respectively. The comparison between the critical exponents for $k$-core percolation and the CF model is presented in section IV. 

\section{$k$-core subgraph and avalanches}

$k$-Core of a network is a subgraph in which the degree of each node is at least $k$. To obtain a $k$-core subgraph, once an ER network of size $N$ with mean degree $z$ is generated, all nodes with degree less than $k$ are deleted. This deletion may decrease the degrees of the remaining nodes. If the degrees of some nodes become less than $k$, then those nodes are deleted as well. This pruning process is repeated until no more node with degree less than $k$ remain in the system. The fraction of nodes remaining in the largest $k$-core subgraph is defined as the order parameter $m$ and the mean degree $z$ is as the control parameter. The order parameter $m$ is large, specifically of $O(1)$ for $z > z_c$, where $z_c$ is a transition point, and decreases continuously with decreasing $z$. As $z$ approaches $z_c$, the deletion of a node from an ER network can lead to the collapse of the giant $k$-core subgraph. Thus, the order parameter is written as follows: 
\begin{equation}
m(q)=\left\{
\begin{array}{lr}
0 & ~{\rm for}~~  z < z_c, \\
m_0+r(z-z_c)^{\beta_m} & ~{\rm for}~~ z \ge z_c, 
\end{array}
\right.
\label{eq:order}
\end{equation}
where $m_0$ and $r$ are constants and $\beta_m$ is the critical exponent of the order parameter. Moreover, the exponent $\gamma_m$ is used to define the fluctuations of the order parameter; that is $\chi_m \equiv N(\langle m^2\rangle -\langle m \rangle^2) \sim (z - z_c)^{-\gamma_m}$. The exponent $\nbar_m$ is used to define the finite size scaling behavior of the order parameter as $m - m_0 \sim N^{-\beta_m/\nbar_m}$ at $z = z_c$. Here the subscript $m$ of the exponents indicates that the exponents are associated with the order parameter $m$.

Next, we remove a randomly selected node from a $k$-core subgraph. This removal may lead to another successive deletion of nodes. The avalanche size is defined as the number of removed nodes in such successive pruning processes and the duration time of the avalanche is the number of pruning steps. These definitions are consistent with those defined in Refs. \cite{baxter,mcc_lee} for the CF model. In the vicinity of $z_c$, an avalanche can lead to the entire collapse of a $k$-core subgraph. We call such an avalanche an infinite avalanche. On the other hand, if the $k$-core still remains after an avalanche, we call it a finite avalanche. 

The size distribution of finite avalanches follows the power law $p_s(z) \sim s_a^{-\tau_a} f(s_a / s_a^*)$, where $s_a$ denotes avalanche size and $f$ is a scaling function. Here $s_a^*$ is the characteristic size of avalanches, which behaves according to $s_a^* \sim (z - z_c)^{-1/\sigma_a}$ as $N \to \infty$ and $s_a^* \sim N^{1/\sigma_a \nbar_a}$ at $z_c$ in finite systems. The exponent $\gamma_a$ is associated with the mean avalanche size as $\langle s_a \rangle \sim (z - z_c)^{-\gamma_a}$. Here the subscript $a$ indicates that the exponents are associated with avalanches. 

In short, we defined two sets of exponents $\{\beta_m, \gamma_m, \nbar_m \}$ and $\{\tau_a, \sigma_a, \gamma_a, \nbar_a \}$ associated with the order parameter and finite avalanches, respectively. These are the same sets already defined in  Ref~\cite{mcc_lee} for the CF model. We also showed that the two sets of exponents are not totally independent, but are related through a new scaling relation $\gamma_a = 1 - \beta_m$. This relation is also valid for the $k$-core problem.  Moreover, we showed that the hyperscaling scaling relation $\nbar_m = 2 \beta_m + \gamma_m$ holds but $\sigma_a \nbar_a = \tau_a$ does not hold. In this paper we show that the results also hold for the $k$-core problem. This suggests that the results may be universal for the hybrid percolation transition induced by the cascading process. 

\section{Results}\label{sec:num_result_ER}

In this section, we show simulation results for ($k=3$)-core percolation. We expect that the obtained results are also valid for general $k$. To confirm our expectation, we tested the critical exponents for the case $k=5$ and found that the critical exponent values are the same within the error as for the case $k=3$. 

\subsection{Critical behavior of the order parameter}
\label{ss:er_crit_order_parameter}

The numerical value of the transition point $z_c$ was estimated in Ref.~\cite{kcore_prx} to be $z_c=3.35091887(8)$, which we confirmed is very precise; thus we adopted this value for our numerical simulations. We numerically calculated the size of discontinuity of the order parameter $m_0=0.267581(6)$ following Ref.~\cite{kcore2}. We analyze our simulation data on the basis of these values. 

\begin{figure}[t]
\includegraphics[width=.95\linewidth]{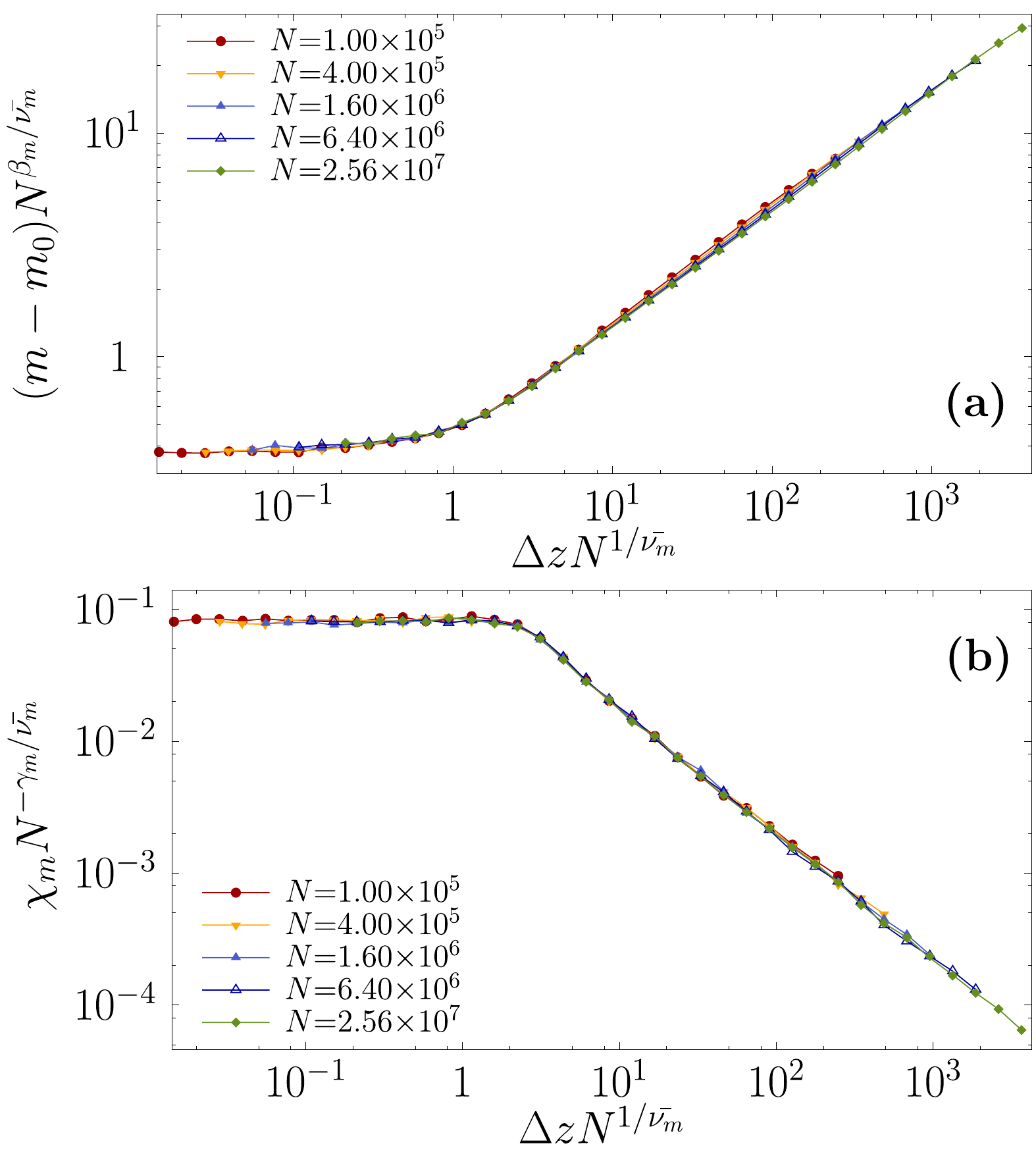}
\caption{
(Color online) (a) Scaling plot of the rescaled order parameter $(m-m_0)N^{\beta_m/\nbar_m}$ versus  $\Delta z N^{1/\nbar_m}$. The data are well collapsed onto a single curve with $\beta_m=0.5$ and $\nbar_m=2.06$.   (b) Scaling plot of $(\langle m^2 \rangle - \langle m \rangle^2)N^{1-\gamma_m/\nbar_m}$ 
for different sizes $N$ versus $\Delta z N^{1/\nbar_m}$, where $\gamma_m=0.97$.}
\label{fig:order}
\end{figure}

\begin{figure}[t]
\includegraphics[width=.95\linewidth]{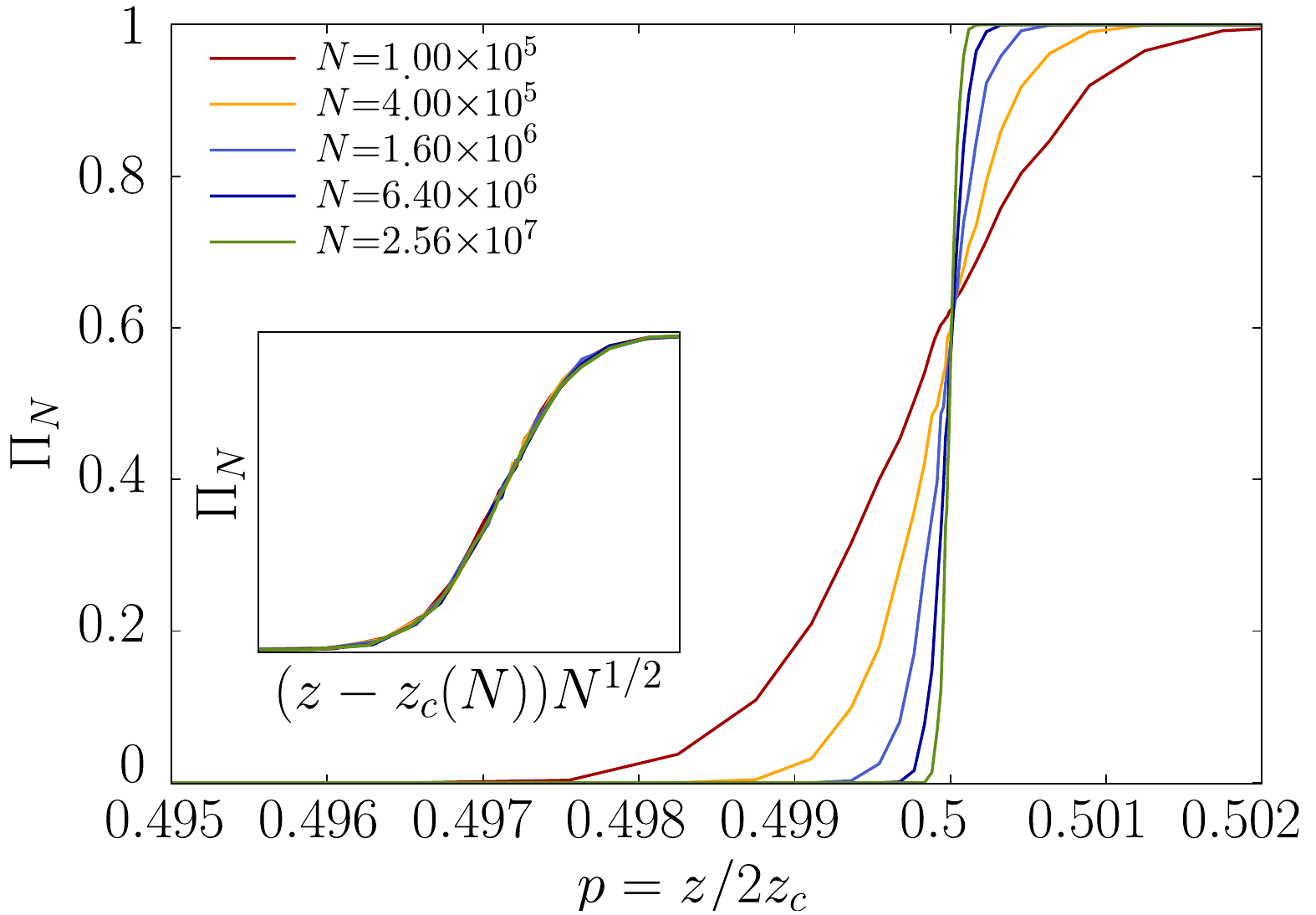}
\caption{
(Color online) The probability $\Pi_N(p)$ that a finite $k$-core subgraph is formed as a function of $p$, where $p = z / 2 z_c$. This form of $p$ easily corresponds to the occupation probability of ordinary ER percolation with mean degree $2z_c$. The critical point $p_c = 0.5$ corresponds to $z = z_c$ in this convention. The inset is a scaling plot in the form of $\Pi_N(z)$ versus \ $(z - z_c(N)) N^{1/2}$. Note that the data collapse is achieved by using $z_c(N)$ instead of $z_c(\infty)$, which is unconventional.}
\label{fig:RN}
\end{figure}

\begin{figure}[h!]
\includegraphics[width=.95\linewidth]{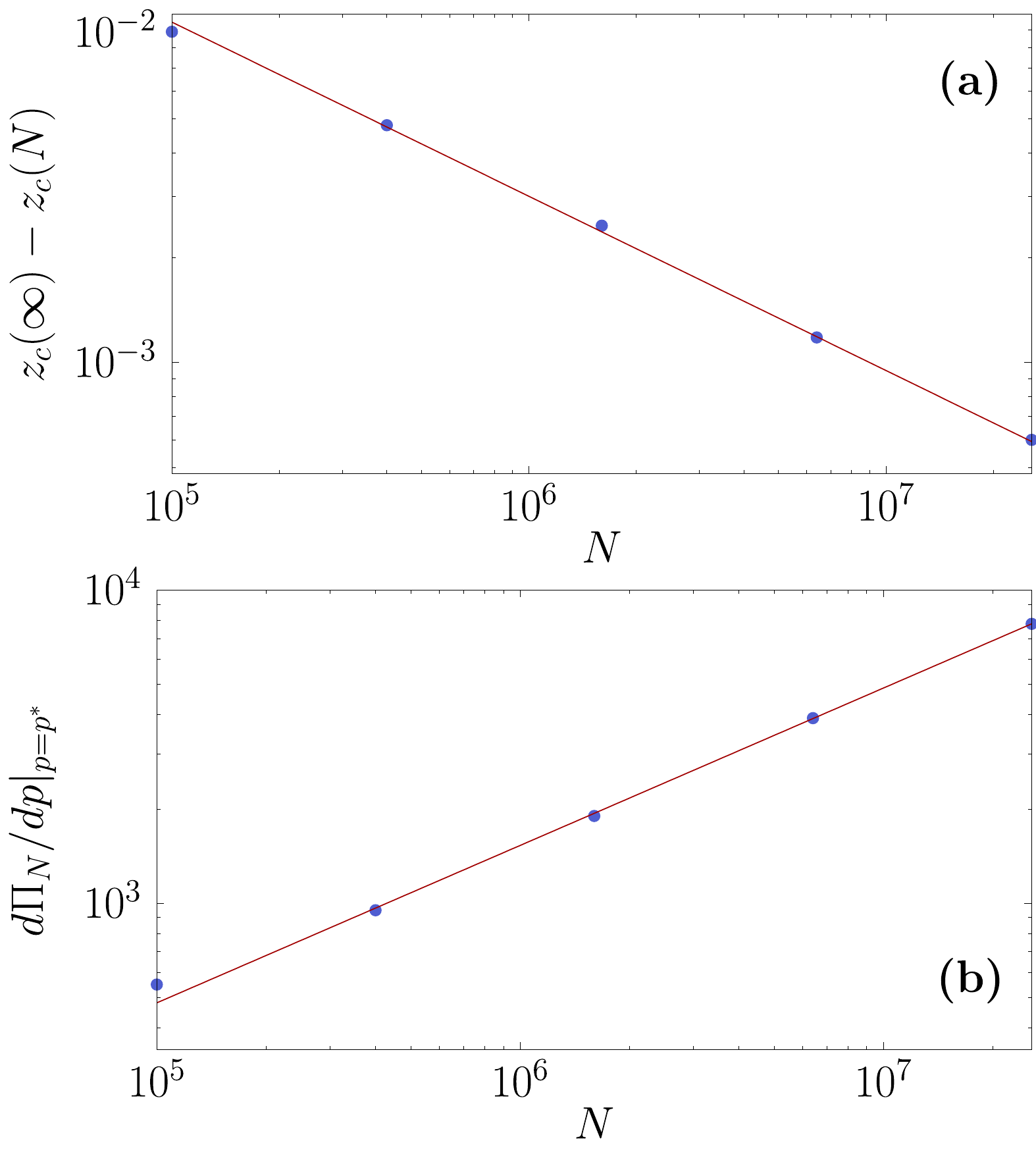}
\caption{(Color online) (a) $z_c(N)$ is the mean value of transition points obtained from different configurations, where the order parameter drops suddenly and $N$ is the system size. It approaches to $z_c(\infty) = 3.35091887\cdots$ as $N$ increases in scale according to $N^{-0.5}$. (b) The slope of $\Pi_N(p)$ at the fixed point $p^*$, at which $p^* = \Pi_N(p^*)$, as a function of the system size $N$. The estimated slope is $0.49 \pm 0.01$; thus, $\nbar^* \approx 2.06\pm 0.04$. }
\label{fig:fssfig}
\end{figure}

We confirm that Eq.~(\ref{eq:order}) is consistent with the theoretical value $\beta_m=1/2$~\cite{kcore1,kcore3}. In Fig.\ref{fig:order}(a), we plot $(m-m_0)N^{\beta_m/\nbar_m}$ versus $\Delta zN^{1/\nbar_m}$ for different system sizes of $N$, where $\Delta z\equiv z-z_c(\infty)$. This confirms the value of $\beta_m$. We also obtain the correlation size exponent $\nbar_m$ as $2.06 \pm 0.05$ using finite size scaling analysis in Fig.~\ref{fig:order}(a). The susceptibility $\chi_m(z)$, the fluctuations of the order parameter, exhibits the divergent behavior $\chi_m \sim (z-z_c)^{-\gamma_m}$ for $z > z_c$. In Fig.~\ref{fig:order}(b), we plot the rescaled quantity $(\langle m^2 \rangle-\langle m \rangle^2)N^{1-\gamma_m/\nbar_m}$ versus $\Delta z N^{1/\nbar_m}$. In the critical $\Delta z$ region, it decays in a power-law manner with exponent $\gamma_m \approx 0.97\pm 0.01$. Moreover, with the choice of $\nbar_m = 2.06$, the data are collapsed well onto a single curve. The obtained exponents $\beta_m\approx 0.5\pm 0.01$, $\gamma_m\approx 0.97\pm 0.01$ and $\nbar_m \approx 2.06 \pm 0.05$ marginally satisfy the hyperscaling relation $\nbar_m=2\beta_m+\gamma_m$.

The probability that a $k$-core subgraph is formed at a certain point $z$, denoted by $\Pi_N(z)$, is also a quantity of interest. $\Pi_N(z)$ approaches a step function in the form of $\Pi_N([z-z_c(N)]N^{1/2})$ as $N$ increases ~(see Fig \ref{fig:RN}). Here $z_c(N)$ is defined as the $z$ intercept of the tangent of $\Pi_N(z)$ at the point where $d\Pi_N(z)/dz$ is the maximum. It behaves as $z_c(\infty) - z_c(N) \sim  N^{-1/\bar{\nu}^*}$ with $1/\bar{\nu}^* \approx 0.49$~(Fig. \ref{fig:fssfig}(a)). This suggests that $\nbar^* \approx 2.06$ ~\cite{stauffer}, which is in accordance with the value $\nbar_m\approx 2.06$ we obtained earlier in Fig.~\ref{fig:order}.

We use the probability $\Pi_N(z)$ for large cell renormalization group transformation~\cite{RSK,herrmann_rg}. To define a coarse-graining procedure, we rescale the control parameter as $p=z/2 z_c$. Then $p$ can be interpreted as the occupation probability of nodes in an ER network with mean degree $2 z_c$, and $\Pi_N(p)$ can be interpreted as the probability that a node is occupied in  a coarse-grained system scaled by $N$. We find the fixed point $p^*(N)$ satisfying $p^*=\Pi_N(p^*)$ and consider the slope $\lambda=\mathrm{d}\Pi_N(p)/\mathrm{d}p$ at $p^*(N)$. We then obtain $\nbar_m=\ln N/\ln \lambda$. Numerically, we find $\lambda \sim N^{0.49\pm 0.01}$; thus $\nbar_m$ is estimated to be $\nbar_m\approx 2.06\pm 0.04$~(Fig. \ref{fig:fssfig}(b)). This value is reasonably consistent with the previous values obtained in this section.

\subsection{Critical behavior of the avalanche size}

\begin{figure}[t]
\includegraphics[width=.95\linewidth]{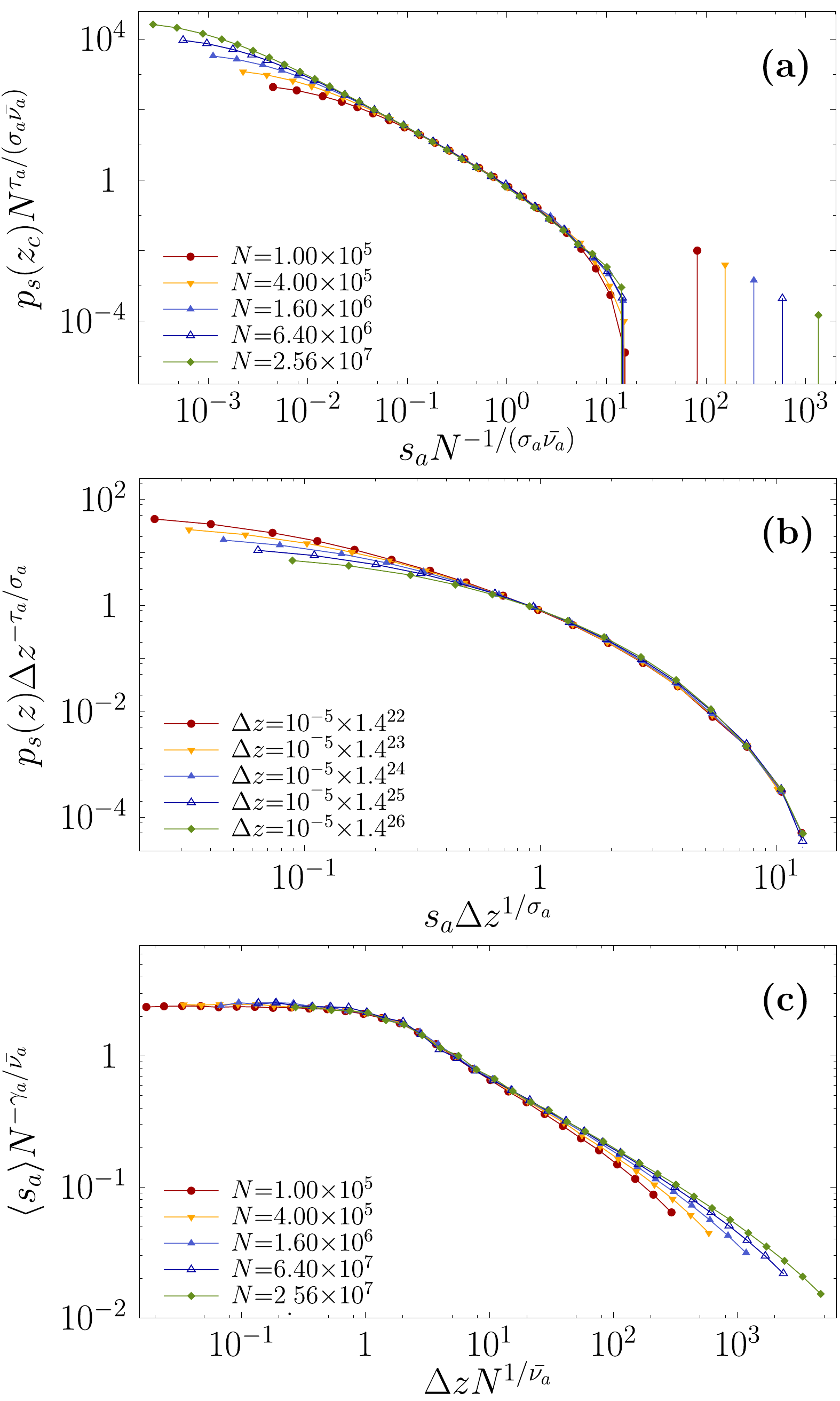}
\caption{(Color online) 
(a) Scaling plot of $p_s(z_c)N^{\tau_a/\sigma_a \nbara}$ versus $s_a/N^{1/\sigma_a\nbara}$ for different system sizes, with $\tau_a=1.5$ and $\sigma_a \nbara = 2.0$. Note that infinite avalanche sizes for different $N$ do not collapse onto a single dot because they depend on $N$ as $s_{a,\infty}^* \sim N$. 
(b) Scaling plot of $p_s(z)\Delta z^{-\tau_a/\sigma_a}$ versus $s_a\Delta z^{1/\sigma_a}$ for different $\Delta z$; here, we consider a fixed system size $N = 2.56\times10^7$, with $\tau_a=1.5$ and $\sigma_a = 1.0$.
(c) Scaling plot of $\langle s_a \rangle N^{-\gamma_a/\nbara}$ versus $\Delta z N^{1/\nbara}$ for different system sizes with $\gamma_a=0.5$.
}
\label{fig:aval}
\end{figure}

The size distribution of finite avalanches behaves similarly to the size distribution of corona clusters. Here a corona cluster is a subgraph of the $k$-core in which the degree of each node is exactly $k$~\cite{kcore_corona}. If we remove a node from a corona cluster, all the nodes in that corona cluster are removed from the $k$-core. Even if a deleted node has degree larger than $k$, if it is connected to one or more nodes belonging to more than one corona cluster, then one of the corona cluster is removed in avalanche processes. However, numerically, in most cases only nodes belonging to one corona cluster are removed and the number of removed nodes with degree larger than $k$ are few when the avalanche size is finite. Thus the statistical feature of finite avalanche sizes and the number of nodes deleted from corona clusters are effectively consistent to each other. Note that in Ref~\cite{kcore_prx}, structure feature of the corona clusters was studied and $\tau_a=3/2$ was analytically obtained using the local tree approximation. We confirm this exponent value in Fig.~\ref{fig:aval}(a) from the size distribution of finite avalanches. 

Avalanches near the transition point need to be classified as finite or infinite avalanches: the former are located separately from the latter as shown in Fig.~\ref{fig:aval}(a). An infinite avalanche is an avalanche whose size per node is as large as $m(z)$. Thus, when this occurs,  the order parameter, the size of the $k$-core per node, completely drops to zero. A universal mechanism underlying such infinite avalanches in the HPTs  of the percolation models such as $k$-core percolation and the CF model on multiplex networks, was recently studied in~\cite{universal_mechanism}.

Fig.~\ref{fig:aval}(a) shows the scaling behavior of the size distribution of finite avalanches in the form of $\psa N^{\tau_a/\sigma_a \nbara}$ versus  $s_aN^{-1/\sigma_a \nbara}$ at $z_c$. The data from different system sizes are well collapsed onto a single curve with the choices of $\tau_a=3/2$ and $\sigma_a\nbara\approx 2.0 \pm 0.03$. This result suggests that there exists a characteristic size  $s_a^*\sim N^{1/\sigma_a \nbara}$ with $\sigma_a\nbara\approx 2.0 \pm 0.03$ for finite avalanches. 
On the other hand, for infinite avalanches, $s_{a,\infty}^*\sim O(N)$. 
For $z > z_c$, we examine the size distribution of finite avalanches for different $\Delta z$ and find that it behaves as $\psa \sim s_a^{-\tau_a} f(s_a/s_a^*)$ where $f(x)$ is a scaling function. Following conventional percolation theory~\cite{stauffer}, we assume $s_a^{*}\sim \Delta z^{-1/\sigma_a}$. The exponent $\sigma_a$ is obtained from the scaling plot of $\psa(z)\Delta z^{-\tau_a/\sigma_a}$ versus $s_a\Delta z^{1/\sigma_a}$  in Fig.~\ref{fig:aval}(b). The data are well collapsed with $\sigma_a\approx 1.0 \pm 0.01$, leading to $\nbar_a\approx 2.0$, which is the same as $\nbar_m$. The numerical value $\sigma_a$ is consistent with the analytic result obtained in Ref.~\cite{kcore3}.

\begin{figure}[t!]
\includegraphics[width=.95\linewidth]{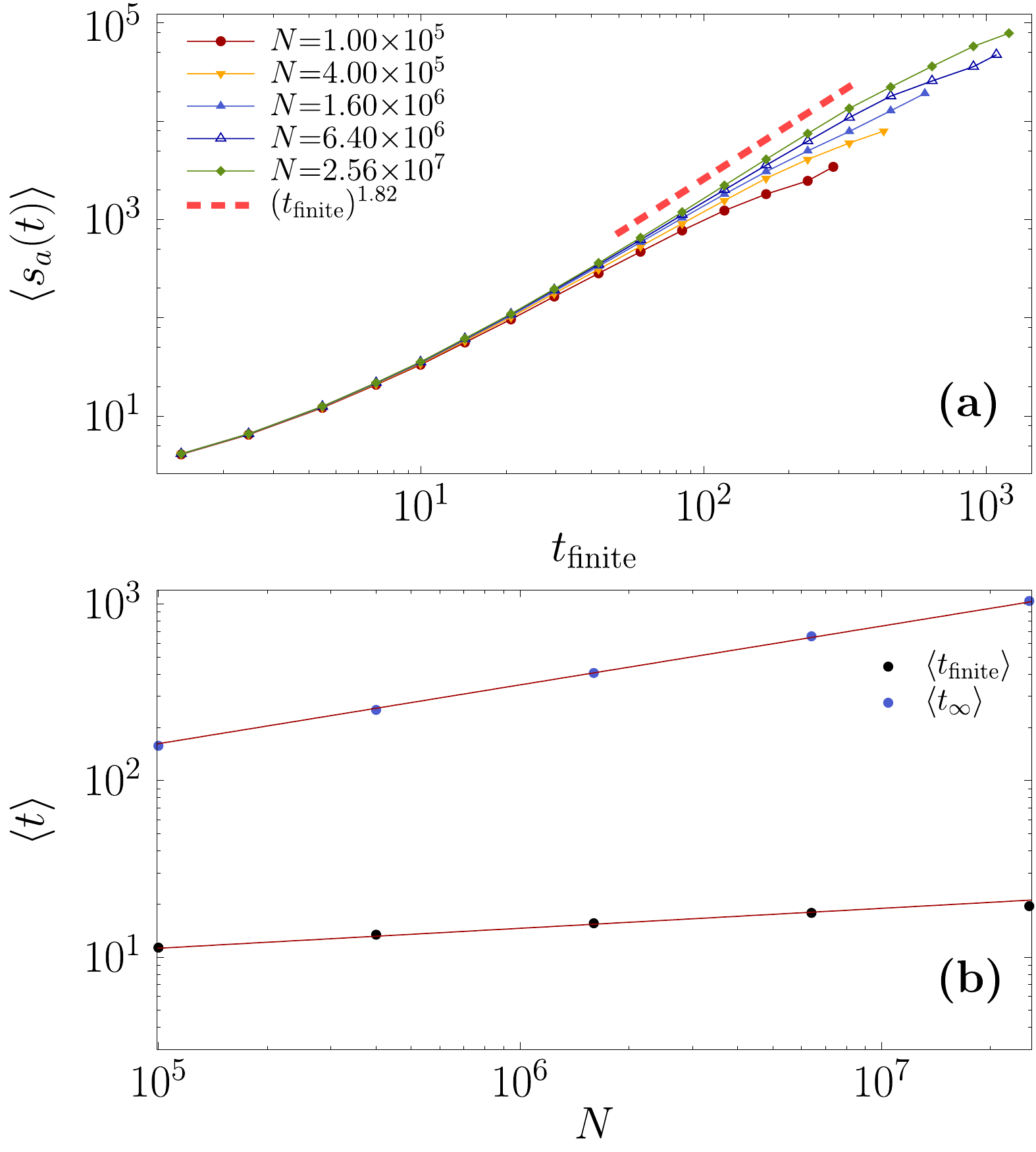}
\caption{(Color online) 
(a) Plot of $\langle s_a(t) \rangle$ as a function of $t$ at $z_c$ for finite avalanches, showing
$\langle s_a(t) \rangle \sim t^{1.82\pm 0.01}$. Dashed line is a guideline with slope 1.82. 
(b) Mean duration time $\langle t_{\rm finite} \rangle$ of finite avalanches and $\langle t_{\infty} \rangle$ of infinite avalanches as a function of $N$ at $z_c$. The slope is $0.10$ for $\langle t_{\rm finite} \rangle$ and $0.33$ for $\langle t_{\infty} \rangle$.
}
\label{fig:hop}
\end{figure}

We examine the mean finite avalanche size $\langle s_a\rangle\equiv \sum_{s_a=1}^{\prime} s_a\psa(z)\sim \Delta z^{-\gamma_a}$, where the prime indicates summation over finite avalanches. It follows that $\gamma_a=(2-\tau_a)/\sigma_a$; thus, $\gamma_a=0.5$ is expected. Our simulation confirms this value in the large-$\Delta z$ region (Fig.~\ref{fig:aval}(c)). Data from different system sizes are well collapsed onto the curve of $\langle s_a \rangle N^{-\gamma_a/\nbara}$ versus $\Delta z N^{1/\nbara}$ with $\gamma_a=0.5$ and $\nbar_a=2.0$.  

Numerically, we determined that the scaling relation $\gamma_a = 1 - \beta_m$ holds as it does for the CF model~\cite{mcc_lee}. Thus, the two sets of exponents are not completely independent. The above scaling relation was derived in Ref.~\cite{mcc_lee} using the simple argument that the mean avalanche size is of the same order as the decrement of the order parameter. 

\subsection{Statistics of the avalanche duration time}
\label{sec:hop-number}

Let $\langle s_a(t) \rangle$ be the mean number of nodes removed accumulated up to the pruning step $t$. In Fig.~\ref{fig:hop}(a), notice  that we obtain $\langle s_a(t) \rangle \sim t^D$ where $D \approx 1.82$.  However, we expect that the fractal dimension converges to $D=2$ asymptotically as $t\to \infty$ when the system size $N\to \infty$. This expectation is based on the following: the nodes removed during the avalanche dynamics are connected and can be regarded as a critical branching tree with mean branching ratio of one~\cite{universal_mechanism}. In this case, the fractal dimension of the critical branching tree is theoretically known to be   $D=2$~\cite{harris, dslee}.  To check the validity of our expectation, we construct a critical branching tree following the idea in Ref.~\cite{kcore_prx}: Each node (ancestor) generates at most two offspring because the nodes are deleted from a $(k=3)$-core subgraph. These numbers are determined stochastically as follows: the probabilities of generating zero, one and two offspring are given by $q_0=1/4$, $q_1=1/2$ and $q_2=1/4$, respectively. Then the conditions $\sum_{i=0}^2 q_i=1$ and $\sum_{i=0}^2 i q_i=1$ to create a critical branching tree of $(k=3)$-core percolation are satisfied. We measure the fractal dimensions of the constructed branching trees from two datasets:  i) one data set composed of tree sizes less than $10^5$ and ii) the other data set composed of unbounded tree sizes. In the dataset i), the tree size limit was imposed to compare it with the data of finite avalanches in $k$-core subgraphs, which have a maximum size $O(10^4)$ for $N=2.56 \times 10^7$. Remarkably, we obtain the fractal dimension $D\approx 1.82$ from data set i), whereas we obtain $D\approx 2.0$ from data set ii). Therefore, the deviation of the numerical value $D\approx 1.82$ from the theoretical value $D=2.0$ is due to the finite-size effect of the simulations.  

\begin{figure}[h]
\includegraphics[width=.95\linewidth]{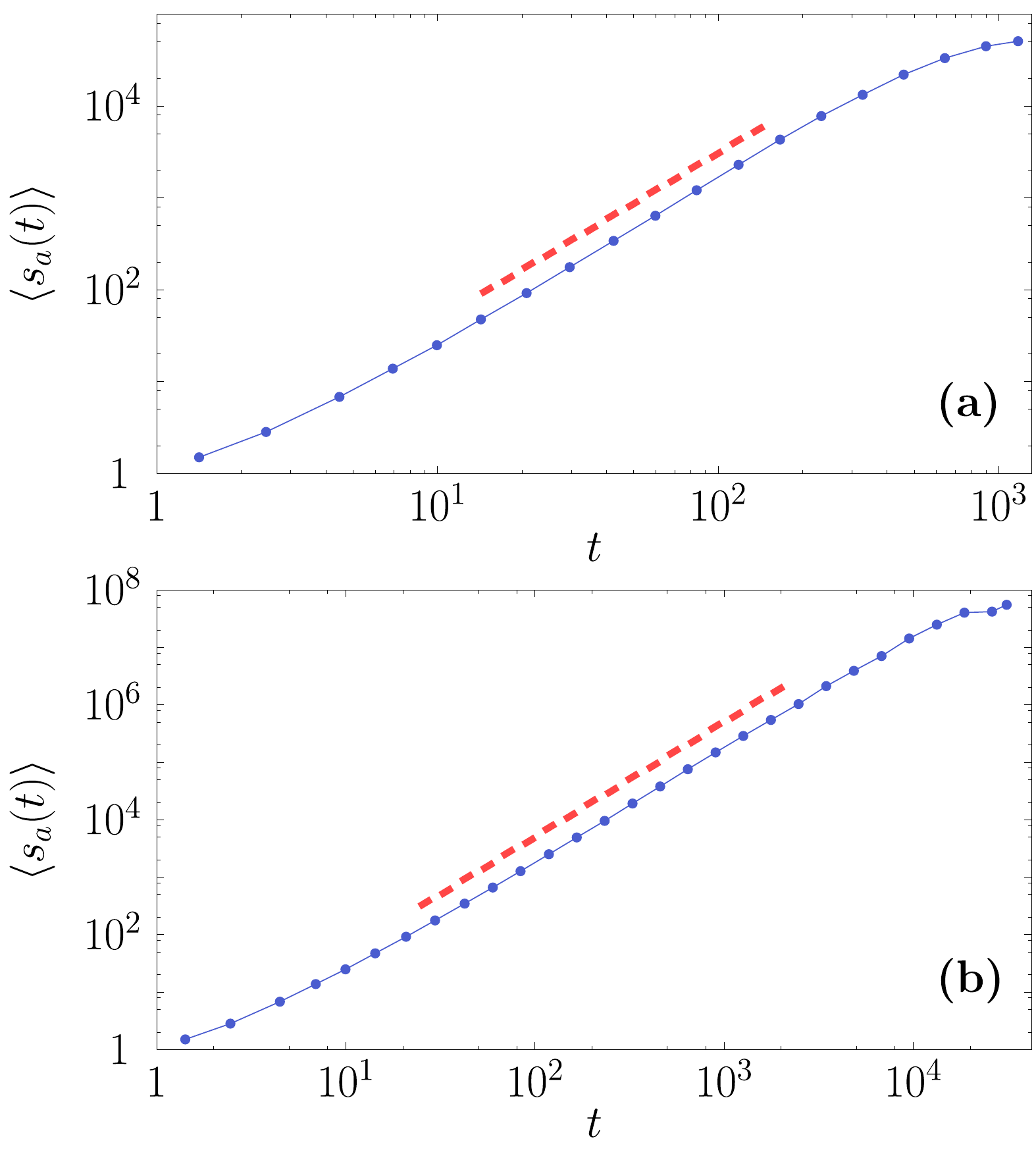}
\caption{(Color online)
Plot of the mean size of critical branching trees $\langle s_a(t)\rangle$ generated by stochastic processes versus the duration time $t$. The data are accumulated from the branching trees of sizes (a) less than $N=10^5$ and unbounded. Dashed lines in (a) and (b) are guidelines with slopes 1.82 and 2, respectively.
}
\label{fig:hopscaling}
\end{figure}

Using the \asd~$p_s(z)$, we set up the duration time distribution $p_t(z)$ through the relations $p_s \mathrm{d}s =p_t \mathrm{d}t$ and $s_a \sim t^{D}$ with $D=1.82$ as $p_{t}(z) \sim t^{-D\tau_a+D-1} f(t^{D}/(\Delta z)^{-1/\sigma_a})$. The scaling plot of $p_t(z)(z-z_c)^{(-D\tau_a+D-1)/D\sigma_a}$ versus $t(z-z_c)^{1/D\sigma_a}$ shown in Fig.~\ref{fig:hopscaling}(a) numerically confirms this. 
The average duration time of finite avalanches $\langle t_{\rm finite} \rangle \equiv \sum_{t=1}^{\prime} t p_t(z)$ follows $\langle t_{\rm finite} \rangle \sim (\Delta z)^{-0.2}$ 
for $z > z_c$ and $\langle t_{\rm finite} \rangle \sim N^{0.10}$ at $z=z_c$ (Figs.~\ref{fig:hop}(b) and \ref{fig:hopscaling}(b).) On the other hand, if we use the theoretical value $D=2$, then the mean avalanche step is obtained as $\langle t_{\rm finite} \rangle\sim \ln N$. 

The mean duration time $\langle t_{\infty} \rangle$ of infinite avalanches scales as $N^{1/3}$, as shown in Fig.~\ref{fig:hop}(b). This result is in agreement with the previous result in Ref.~\cite{kcore_attack}. This scaling is universal for infinite avalanches in the HPTs  of percolation models~\cite{universal_mechanism}.

\begin{figure}[h]
\includegraphics[width=.95\linewidth]{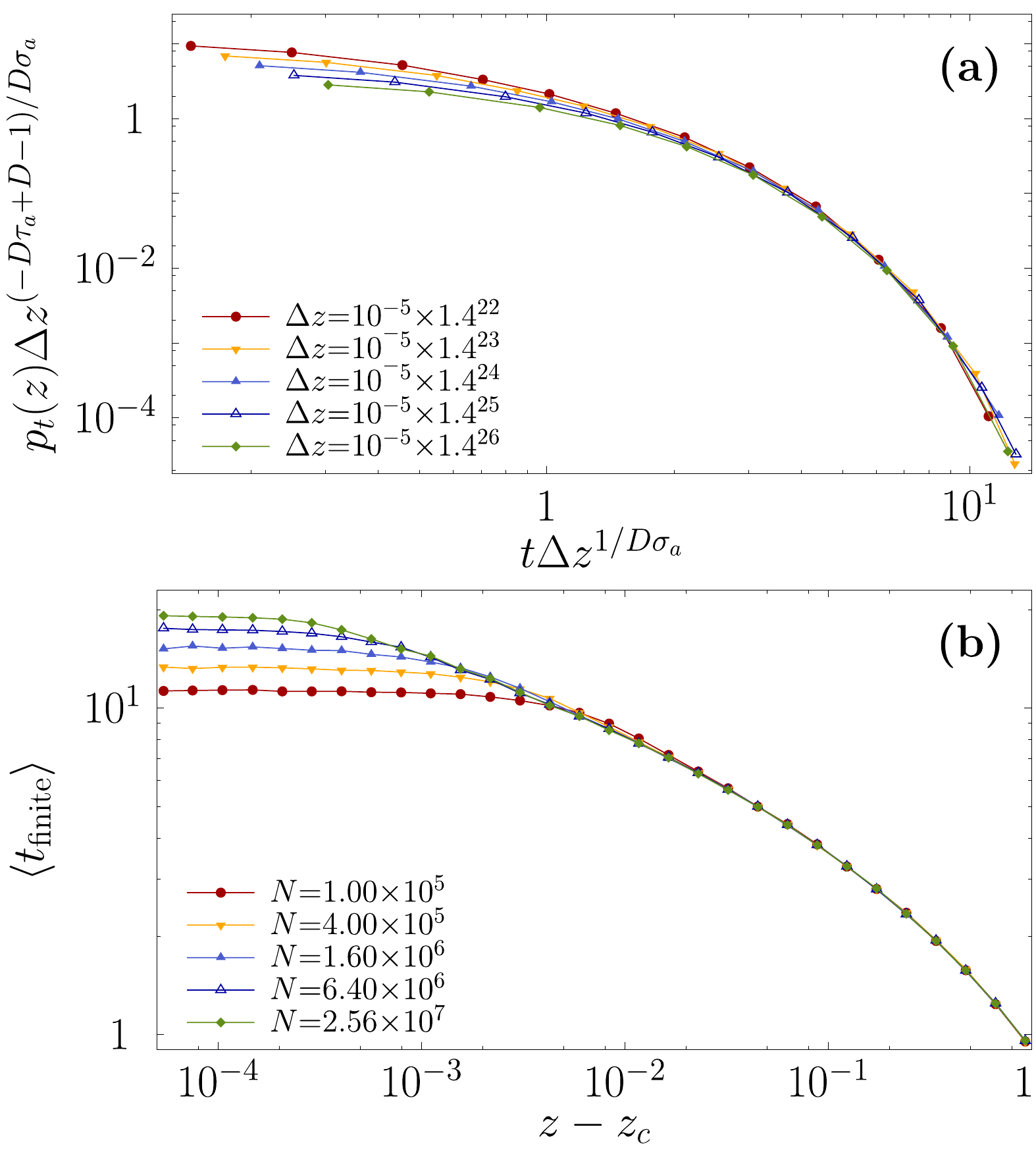}
\caption{(Color online)
(a) Scaling plot of the distribution $p_t$ as a function of the duration time $t$ for finite avalanches with $D = 1.82$.
(b) Plot of the mean duration time of finite avalanches $\left< t_{\text{finite}} \right>$ versus $z-z_c$ for different system sizes.
}
\label{fig:hopscaling}
\end{figure}

\section{Discussions and summary}

We have studied the critical behaviors of the HPT of $k$-core percolation on ER networks. We found that two diverging behaviors occur at a transition point: the fluctuations of the order parameter and the mean size of finite avalanche. These two divergences cannot be represented by single critical exponent $\gamma$.  Thus, we introduced two sets of exponents $\{\beta_m, \gamma_m, \nbar_m \}$ and $\{\tau_a, \sigma_a, \gamma_a, \nbar_a \}$ associated with the order parameter and finite avalanches, respectively. As noted explicitly in Ref.~\cite{mcc_lee}, the two sets of exponents are not completely independent, but are related through a new scaling relation $\gamma_a = 1 - \beta_m$. Moreover, we showed that the hyperscaling scaling relation $\nbar_m = 2 \beta_m + \gamma_m$ holds but $\sigma_a \nbar_a = \tau_a$ does not hold. This feature also appear in the CF model on multiplex networks, thus may be universal for hybrid percolation transitions induced by cascade pruning processes. 

For comparison, we list the numerical values of the critical exponents of each set that we obtained for $k$-core percolation and the CF model in Table I.  Note that the exponents $\beta_m$, $\tau_a$, $\sigma_a$, $\gamma_m$, $\gamma_a$,  ${\bar \nu}_m$ and the fractal dimension $D$ of finite avalanches are consistent with each other for $k$-core percolation and the CF model. However, ${\bar \nu}_a$ and ${\bar \nu}^*$ are different from each other. Furthermore, note that the two exponents ${\bar \nu}_m$ and ${\bar \nu}_a$ are consistent within the error bar for $k$-core percolation, however, they are different for the CF model. Thus, it seems that the exponent ${\bar \nu}_a$ depends on detailed dynamics.    

Finally, note that the critical exponents obtained for finite clusters $\tau_a=3/2$, $\sigma_a \approx 1.0$ and $\gamma_a\approx 0.5$ are consistent with those obtained analytically from corona clusters~\cite{kcore_prx}. Thus, $\gamma_a \approx 0.5$ is different from the exponent $\gamma_m\approx 0.97$ for the fluctuations of the order parameter.  This result is not compatible with the fact that diverging behaviors of the fluctuations of the order parameter and the mean cluster size for ordinary percolation are represented by single critical exponent $\gamma$. 

\begin{table*}
\caption{Numerical values of the critical exponents. Here $m$ and $a$ in the second column represent that the exponents are associated with the order parameter and the avalanche, respectively. Moreover, $\bar{\nu}^*$ is the exponent defined by the relation $z^*(N)-z_c(\infty)\sim N^{-1/\bar{\nu}^*}$, and $D$ is the fractal dimension of the mean size of finite avalanches, defined by $\langle s_a(t) \rangle \sim t^D$. Note that the hyperscaling relation $2\beta_m+\gamma_m={\bar \nu_m}$ is satisfied. ** is the theoretical fractal dimension of a critical branching tree.}
\begin{center}
\setlength{\tabcolsep}{12pt}
{\renewcommand{\arraystretch}{1.5}
\begin{tabular}{ccccccccc}
    \hline
    \hline
    
    & & $\beta$ & $\tau$ & $\sigma$ & $\gamma$ & $\bar{\nu}$ & $\bar{\nu}^*$ & $D$ \\
    \hline
    
    \multirow{2}{*}{$k$-core} & $m$ & $0.5 \pm 0.01$ & - & - & $0.97 \pm 0.01$ & $2.06 \pm 0.05$ & \multirow{2}{*}{$2.0$} & -\\
    \cline{2-2}
    & $a$ & - & $1.5 \pm 0.01$ & $1.0 \pm 0.01$ & $0.52 \pm 0.01$ & $2.0 \pm 0.01$ & & $1.82 \pm 0.02$ \\
    & & & & & & & & ($D=2$)$^{**}$\\
    \\
    \multirow{2}{*}{CF model}
    & $m$ & $0.5 \pm 0.01$ & - & - & $1.05 \pm 0.05$ & $2.1 \pm 0.02$ & \multirow{2}{*}{$1.5$} & -\\
    \cline{2-2}
    & $a$ & - & $1.5 \pm 0.01$ & $1.0 \pm 0.01$ & $0.5 \pm 0.01$ & $1.85 \pm 0.02$ & & $2.0 \pm 0.01$ \\
    
    \hline
    \hline
\end{tabular}}
\label{table}
\end{center}
\end{table*}

\begin{acknowledgments}
This work was supported by the National Research Foundation in Korea with grant No.\ NRF-2014R1A3A2069005. D.L. and M. J.  contributed equally to this work. 
\end{acknowledgments}


\end{document}